\documentclass[aps,preprint,nofootinbib,unsortedaddress,floatfix,superscriptaddress]{revtex4}
\usepackage{epsfig}
\usepackage{graphicx}
\usepackage{multirow}
\usepackage{latexsym,amsbsy,amsmath}
\usepackage{soul,xcolor}
\usepackage{stackrel}
\begin{document}
\title{Pauli principle forbids $\Omega_{QQQ}\Omega_{QQQ}\Omega_{QQQ}$ bound states}
\author{H.~Garcilazo} 
\email{hgarcilazos@ipn.mx} 
\affiliation{Escuela Superior de F\' \i sica y Matem\'aticas, \\ 
Instituto Polit\'ecnico Nacional, Edificio 9, 
07738 Mexico D.F., Mexico} 
\author{A.~Valcarce} 
\email{valcarce@usal.es} 
\affiliation{Departamento de F\'\i sica Fundamental,\\ 
Universidad de Salamanca, E-37008 Salamanca, Spain}
\date{\today} 

\begin{abstract}
Lattice QCD studies have shown the attractive character of the 
$^1S_0$ $\Omega_{QQQ}\Omega_{QQQ}$, $Q=s,c,b$, interaction,
predicting deeply
bound states as the mass of the heavy quark increases. 
This has led to the question of the possible existence of bound 
states of more than two $\Omega_{QQQ}$ baryons, in particular
$\Omega_{QQQ}\Omega_{QQQ}\Omega_{QQQ}$ bound states.
We discuss how these states might not exist in nature in 
any flavor sector. The reason would be due to the simultaneous 
action of two consequences 
of the Pauli principle in the different partial waves:
one at the baryon level, affecting the $^1S_0$ partial wave, 
and the other due to the quark substructure, 
affecting the $^5S_2$ partial wave.
\end{abstract} 
\maketitle

\section{Introduction.}
\label{secI}
Experimental findings since 2003 with the discovery of the $X(3872)$~\cite{Cho03}
have opened a new era for hadron spectroscopy. 
As a general conclusion, the idea has emerged that the heavy hadron spectra 
shows the contribution of states that do not belong to the simplest 
quark-antiquark (meson) or three-quark (baryon) structures proposed 
by Gell-Mann~\cite{Gel64}. Experimental evidence of more complex 
structures is steadily and continuously being found~\cite{Hay24}.

One of the most striking predictions of the quark model
was the possible existence of bound states in double heavy 
tetraquark configurations, $QQ\bar q\bar q$~\cite{Ade82}.
Stability was shown to be a function of the mass ratio $M_Q/m_q$. 
The mechanism that stabilizes $QQ\bar q\bar q$ at large $M_Q/m_q$ 
is the same that makes the hydrogen molecule much more stable than the 
positronium one in atomic physics~\cite{Ric93}. 
Such prediction was confirmed by the LHCb Collaboration that announced the 
discovery of a very sharp peak in the $DD\pi$ spectrum that was dubbed 
$T_{cc}$~\cite{Aai22,Aaj22}. It corresponds to a minimal quark content 
$cc\bar u\bar d = T_{cc}^+$. The $T_{cc}^+$ misses binding by a 
very small amount. So, it is almost certain that the heavier partners, 
$T_{bc}=bc \bar u\bar d$ and $T_{bb}=bb\bar u\bar d$, could be stable 
with respect to the strong and electromagnetic interactions~\cite{Aai22,Col24}.

Another area that has recently attracted a great deal of interest is the case
of dibaryons in the heavy flavor sectors. Only one stable state made of two 
baryons is known, the deuteron. This is a loosely bound state of a proton and 
a neutron with spin 1. Lattice QCD studies have
opened the door to the existence of stable states made of heavy flavor
baryons. Thus, the HAL QCD Collaboration~\cite{Gon18} has studied the 
$\Omega \Omega$ system in the $^1S_0$ channel through $(2+1)-$flavor lattice
QCD simulations with a large volume and nearly physical pion mass 
$m_\pi \simeq 146$ MeV. Their results indicate that the $^1S_0$ $\Omega\Omega$ 
state has an overall attraction and is located near to the unitary regime
with a binding energy
$B^{\rm QCD}_{\Omega\Omega}=1.6(6)\left(^{+0.7}_{-0.6}\right)$ MeV.
The Coulomb repulsion reduces the binding energy by a factor of two,
$B^{\rm QCD+Coulomb}_{\Omega\Omega}=0.75(5)(5)$ MeV, the system being still
bound.

A few years later, the $\Omega_{ccc}\Omega_{ccc}$ system was studied
on the basis of the HAL QCD method~\cite{Lyu21}. Without the 
Coulomb interaction, a $(2+1)-$flavor
lattice QCD study of the $^1S_0$ channel with nearly physical light-quark 
masses and the relativistic
heavy-quark action with the physical charm quark mass,
shows a weak repulsion at short distances surrounded by
a relatively strong attractive well, which leads to a 
bound state with a binding energy 
$B^{\rm QCD}_{\Omega_{ccc}\Omega_{ccc}}=5.68(0.77)\left(^{+0.46}_{-1.02}\right)$ MeV.
Taking into account the Coulomb repulsion between the $\Omega^{++}_{ccc}$'s,
with their charge form factor obtained from lattice QCD,
a scattering length of
$a^C_0 = -19(7)\left(^{+7}_{-6}\right)$ fm was obtained.
The ratio $r^C_{\rm eff}/a^C_0 = - 0.024$ is considerably smaller than 
that of the dineutron ($-0.149$), which
indicates that $\Omega_{ccc}\Omega_{ccc}$ is located in the unitary regime.

More recently, the first lattice QCD investigation
of heavy dibaryons $\Omega_{bbb}\Omega_{bbb}$
in the $^1S_0$ channel has been carried out~\cite{Mat23}. 
It was reported a binding energy  
$B^{\rm QCD}_{\Omega_{bbb}\Omega_{bbb}}= 81\left(^{+16}_{-14}\right)$ MeV.
With such a deep binding, Coulomb repulsion serves only as a perturbation 
on the ground state wave function of the parametrized strong potential and 
may shift the strong binding only by a few percent. In particular, 
the associated maximum change in binding
energy is found to be between 5 and 10 MeV. 

On the other hand, it is also well known from nuclear 
physics that when a two-baryon interaction 
is attractive, if such a system is merged with additional nuclear matter
and there are no severe Pauli principle constraints, the attraction 
can be strengthened.
Simple examples can be given of the effect of a third or 
a fourth baryon in two-baryon systems with attractive character interactions. The deuteron, $(I)J^P=(0)1^+$, 
is bound by $2.225$ MeV, while the triton,
$(I)J^P=(1/2)1/2^+$, is bound by $8.480$ MeV, 
and the $\alpha$ particle, $(I)J^P=(0)0^+$,
is bound by $28.295$ MeV. The binding per nucleon $B/A$ increases as $1:3:7$.
A similar argument is found in systems with strangeness $-1$.
The hypertriton $^3_\Lambda$H, $(I)J^P=(0)1/2^+$, is bound with a separation
energy of $130 \pm 50$ keV~\cite{Jur73}\footnote{Note that the above 
canonical value of the hypertriton binding energy has been
challenged recently by the STAR Collaboration~\cite{Ada20}, claiming
a much more tightly bound hypertriton with
$B= 0.41 \pm 0.12 \, {\rm (stat)} \pm 0.11 \, {\rm (syst)}$ MeV, which does not affect our reasoning.}, 
and the $^4_\Lambda$H, $(I)J^P=(0)0^+$, is bound
with a separation energy of 
$2.12 \pm 0.01 \, {\rm (stat)} \, \pm 0.09 \, {\rm (syst)}$ MeV~\cite{Ess15}. 

The template that has been observed for tetraquarks, 
that the binding energy increases for  heavy quarks~\cite{Ade82,Col24}, 
it therefore appears to be reproduced in the case of other multiquark
hadrons involving more than one bottom quark~\cite{Mat23}.
Considering all the above together, a
common interesting pattern might be emerging that the presence
of more than one bottom quark enhances the binding in
multihadron systems~\cite{Kar17,Eic17}.
Given the bound-state nature of two--$\Omega$ systems in the 
different heavy flavor sectors predicted by lattice 
QCD calculations ~\cite{Gon18,Lyu21,Mat23},
the question immediately arises whether systems containing more than two such 
$\Omega$-like baryons might also be bound in nature~\cite{Wul23,Gar24}. 
This issue is particularly
relevant in the bottom sector given the very large binding energy 
predicted for the two-body system by 
lattice QCD calculations~\cite{Mat23}. This is the question we analyze 
in this work making use of the quark substructure of 
the $\Omega^i$ baryons~\footnote{Since our arguments will be
valid for any of the flavor sectors considered, henceforth we will denote
by $\Omega^i$ the $\Omega$ for $i=s$, the $\Omega_{ccc}$ for $i=c$ 
and the $\Omega_{bbb}$ for $i=b$.}, similar to what was 
done in the case of tetraquarks~\cite{Ade82} and other two-baryon
systems~\cite{Gar97,Val05}.

\section{$\Omega^i\Omega^i\Omega^i$ states.}
\label{secII}
Due to the Fermi-Dirac statistic the $\Omega^i\Omega^i$ system can
only exist in two different $L=0$ partial waves: $^1S_0$ and $^5S_2$.
Moreover, there is only one fully antisymmetric $\Omega^i\Omega^i\Omega^i$ state
with three $\Omega^i$'s in relative $S$ wave, the $J^P=3/2^+$.
Both two-body partial waves, $^1S_0$ and $^5S_2$, are basic 
to construct the fully antisymmetric $J^P=3/2^+$ 
$\Omega^i\Omega^i\Omega^i$ state. 

Let's first study
the three-body problem considering only the contribution of the attractive interaction
in the $^1S_0$ $\Omega^i\Omega^i$ channels, which are known 
in detail from the lattice QCD calculations~\cite{Gon18,Lyu21,Mat23}. 
The objective of this first analysis will be to understand
if the attractive two-body interactions, 
especially in the $\Omega^b\Omega^b$ case, are reflected in 
an attractive character of the three-body system or 
if, on the contrary, there are Pauli effects that mask such attraction 
as it occurs in other three-body systems~\cite{Gar87}.

We have solved the $\Omega^i\Omega^i\Omega^i$ three-body problem by the Faddeev
method as described in detail, for example, in Ref.~\cite{Gar97}. We have used
as input the $^1S_0$ $\Omega^i\Omega^i$
potentials parametrized in Refs.~\cite{Gon18,Lyu21} for $i=s$ and $c$. 
In the case of the $i=b$ potential
shown in Fig. 4 of Ref.~\cite{Mat23} we have parametrized the interaction by 
means of three Gaussians as done in Refs.~\cite{Gon18,Lyu21}. 
In spite of the attractive character of the $^1S_0$ two-body interaction,
we have not found any $J^P=3/2^+$ three-body bound state in any 
flavor sector considering only the $^1S_0$ 
$\Omega^i\Omega^i$ channels.

It is important to realize that this result is a direct consequence of 
the Pauli principle at the baryon level.
It is due to the fact that the recoupling
coefficient between the two spin-0 Faddeev amplitudes
is a negative number, see Table~\ref{tab1}, 
so that it effectively changes the nature of the 
two-body interaction from attractive to repulsive such that no bound 
state can be obtained in a one-channel calculation~\cite{Gar87}.
In fact, increasing the attraction in the $^1S_0$ two-body channel 
results in an increase of the repulsion in the three-body system.
This result applies for all systems with three identical 
fermions, in particular for the three-neutron case. However, 
unlike the case of neutrons that have spin $1/2$, 
the $\Omega^i$'s are spin 3/2 fermions and therefore,
as explained above, for $L=0$, in addition to the spin-0 two-body 
amplitudes also contribute spin-2 two-body amplitudes.

If we choose an arbitrary three-body Jacobi coordinate system,
$\left|s_{12},s_3;S\right>$, 
the fully antisymmetric three-body $J^P=3/2^+$ 
wave function has two 
different components, $s_{12}=0$ and $s_{12}=2$, that has to be expressed 
in the different Jacobi couplings, $\left|s_1,s_{23};S\right>$.
The recoupling coefficients between the different Jacobi coordinate
systems are given in Table~\ref{tab1}.
\begin{table}[t]
\begin{ruledtabular}
\begin{tabular}{||cc|ccc||}
&& \multicolumn{2}{c}{$\left|s_{12},s_3;S\right>$}&        \\
  && $s_{12}=0$   &  $s_{12}=2$ &\\ \hline
\multirow{2}{*}{$\left|s_1,s_{23};S\right>$}	&   $s_{23}=0$  &   $-\frac{1}{4}$ & $-\frac{\sqrt{5}}{4}$  & \\
                                              &   $s_{23}=2$  & $-\frac{\sqrt{5}}{4}$ & $\frac{3}{4}$  &\\
\end{tabular}
\caption{Recoupling coefficients between the different Jacobi coordinate systems
of the $\Omega^i\Omega^i\Omega^i$ $J^P=3/2^+$ state.}
\label{tab1}
\end{ruledtabular}
\end{table}
This table illustrates on the one hand the relevance of 
the $^5S_2$ partial wave 
to the $\Omega^i\Omega^i\Omega^i$ $J^P=3/2^+$ state, 
much higher than that of the attractive $^1S_0$ partial wave, and on the other
hand the strong coupling between the $^1S_0$ and the $^5S_2$ partial waves
in the $J^P=3/2^+$ three-body state.
Moreover, the recoupling coefficient between the spin-2 amplitudes is 
positive and therefore free of the Pauli principle effect
at the baryon level discussed above, that effectively changes the nature 
of the spin-0 two-body interaction from attractive to repulsive
in the three-body system. 

Therefore, a definitive conclusion about the existence
of bound states in the only 
$\Omega^i\Omega^i\Omega^i$ $S$-wave state, $J^P=3/2^+$,
cannot be reached without a full fledged 
calculation considering the contribution 
of both two-body partial waves: $^1S_0$ and $^5S_2$.

\section{$^5S_2$ $\Omega^i \Omega^i$ state.}
We have explained in the introduction that lattice QCD simulations have 
derived the $^1S_0$
interaction for the different $\Omega^i\Omega^i$ systems, but so far 
the interaction in the $^5S_2$ channel has not been derived.
A naive ansatz of spin independence of the $\Omega^i\Omega^i$ interaction~\cite{Gar24},
as it could be deduced from a Fermi-Breit like potential at baryonic level, 
leads to bound states of hundreds of MeV for the $i=b$ case using the 
$^1S_0$ lattice QCD interaction of Ref.~\cite{Mat23}.
This would be a surprising result with binding energies even higher 
than expected for the beautiful partners of the $T^+_{cc}$ 
tetraquark~\cite{Kar17,Eic17,Her20} even more so considering that 
if the quark substructure is considered and the six-particle system 
is treated rigorously no bound states are found for very heavy 
flavor dibaryons~\cite{Ric20} nor for fully-heavy tetraquarks~\cite{Ric17}.
Thus, as it has been done for multiquark states~\cite{Ade82,Her20,Ric20,Ric17} 
and other two-baryon systems~\cite{Val05,Gar97,Val97}, 
one can use the quark substructure 
of the $\Omega^i$ baryons to gain some clues about the most
significant features of the $\Omega^i\Omega^i$ interaction
in the $^5S_2$ partial wave.

Consequently, it is important to keep in mind that we are working 
with identical baryons, $\Omega^i$, that are formed by identical quarks, 
which is not always the case for identical hadrons.
Therefore, the antisymmetry of the two-baryon
wave function to the exchange of identical quarks between the 
two baryons must be also guaranteed in the overlap region of the two baryons.  
If there were any consequences due to the identity of the baryon
constituents, it would be reflected in the normalization of the 
wave function that appears in the calculation of any observable, 
in particular the potential as we will discuss in more detail in the next 
section. Thus,
we describe the baryon-baryon system by means of a constituent quark
cluster model, i.e., baryons are described as clusters of three constituent
quarks. Assuming a two-center shell model the wave function of
an arbitrary baryon-baryon system can be written as~\cite{Val05}:
\begin{eqnarray}
\Psi_{B^\alpha B^\beta}^{{LSI}}({\vec R}) & = & {\frac{\cal A}{\sqrt{1 +
\delta_{B^\alpha B^\beta}}}} \sqrt{\frac{1}{2}} \Biggr\{ \left[
\Phi_{B^\alpha} \left( 123 ;{-\frac{\vec R}{2}} \right)
\Phi_{B^\beta} \left( 456 ; {\frac{\vec R}{2}} \right)
\right]_{{LSI}} \, + \nonumber \\
& + &(-1)^{f} \,
\left[
\Phi_{B^\beta} \left( 123 ; {-\frac{\vec R}{2}} \right)
\Phi_{B^\alpha} \left( 456 ; {\frac{\vec R}{2}} \right)
\, \right]_{{LSI}} \Biggr\} \,
,\label{Gor}
\end{eqnarray}
where ${\cal A}$ is the antisymmetrization operator accounting for 
the possible existence of identical quarks inside the hadrons. The symmetry 
factor $f$ satisfies $L+S_1+S_2-S+I_1+I_2-I+f=$ odd. For non-identical baryons $f$
indicates the symmetry associated to a given set of values $LSI$. The non-possible 
symmetries correspond to forbidden states. For identical baryons, $B^\alpha=B^\beta$,
$f$ has to be even in order to have a non-vanishing wave function, recovering 
the well-known selection rule
$L+S+I=$ odd. In the case we are
interested in, two baryons made of identical quarks with $I=0$, the antisymmetrization operator comes given by,
\begin{equation}
{\cal A} = \left ( 1-\sum_{i=1}^3 \sum_{j=4}^6 P^{LSC}_{ij} \right )(1-{\cal P}) \label{ant} \, ,
\end{equation}
where $P^{LSC}_{ij}$ exchanges a pair of identical quarks $i$ and $j$ in the configuration, spin and color spaces, and $\cal P$
exchanges identical baryons.
Assuming a Gaussian form for the wave functions
of the quarks inside the hadrons,
the normalization of the two--baryon wave function 
$\Psi_{B^\alpha B^\beta}^{LS(I=0)}({\vec R})\equiv \Psi_{B^\alpha B^\beta}^{LS}({\vec R})$ of Eq.~(\ref{Gor})
can be expressed as~\cite{Val05,Val97,Oka87},
\begin{figure}[t]
\vspace*{-4.5cm}
\includegraphics[width=.6\columnwidth]{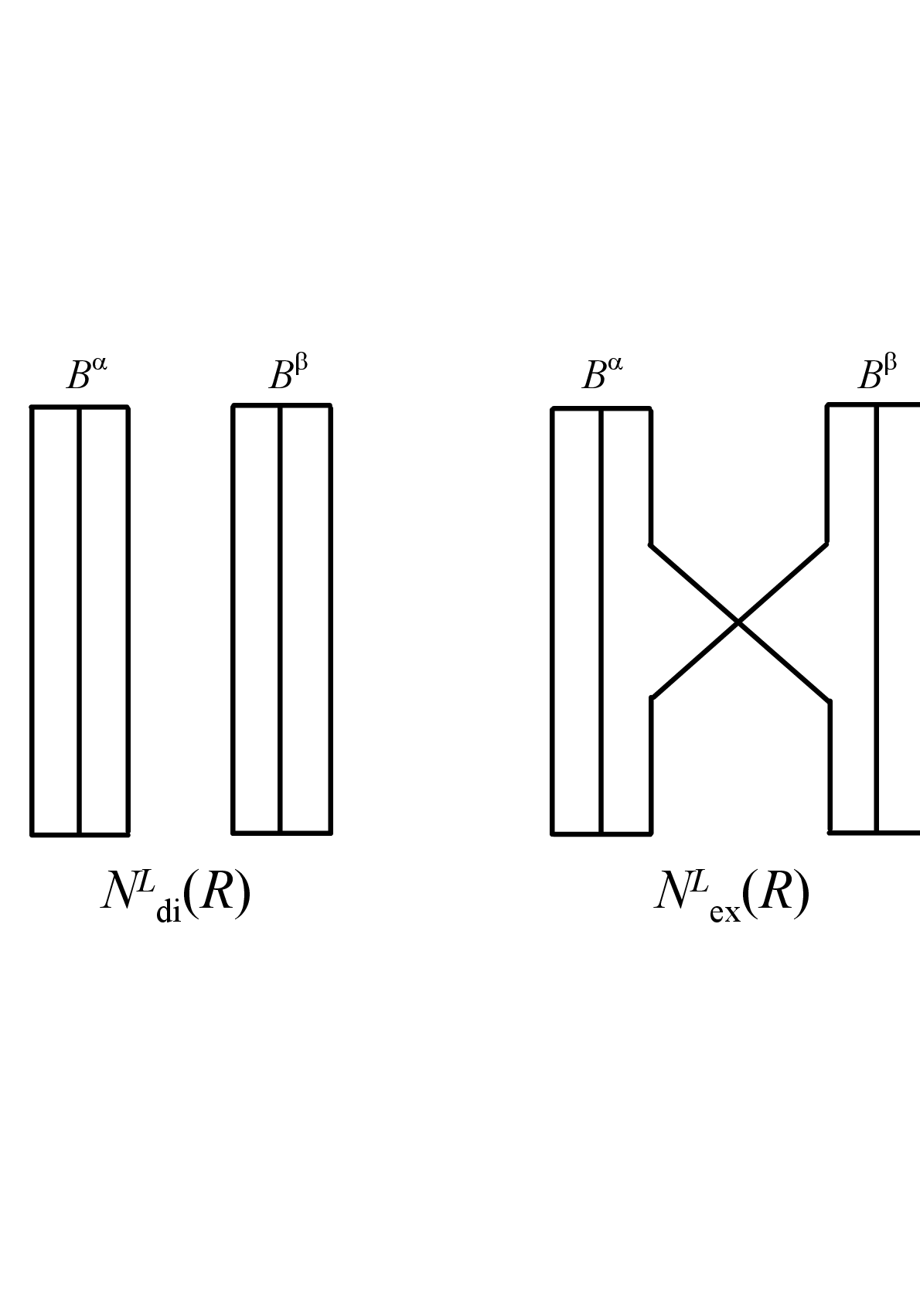}
\vspace*{-4.6cm}
\caption{Normalization diagrams of the two-baryon wave function.}
\label{fig1}
\end{figure}
\begin{equation}
{\cal N}_{B^\alpha B^\beta}^{LS}(R)= 
\left \langle \Psi_{B^\alpha B^\beta }^{LS} ({\vec %
R}) \mid \Psi_{B^\alpha B^\beta }^{LS} ({\vec R}) \right \rangle =
N^{L}_{\rm di}(R) - C(S) \, N^{L}_{\rm ex}(R),
\label{norm}
\end{equation}
where $N^{L}_{\rm di}(R)$ and $N^{L}_{\rm ex}(R)$ are the direct
and exchange radial normalization contributions shown in Fig.~\ref{fig1},
whose explicit expressions are,
\begin{eqnarray}
 N_{\rm di}^{L} (R) &=& 4 \pi \exp \left( -\frac{3}{4} \, \frac{R^2}{\alpha^2} 
\right)  i_{L+1/2} 
\left( \frac{3}{4} \frac{R^2}{\alpha^2} \right) \,, \\ \nonumber
N_{\rm ex}^{L} (R) &=& 4 \pi \exp \left( -\frac{3}{4} \frac{R^2}{\alpha^2} 
\right)  i_{L+1/2} 
\left( \frac{1}{4} \frac{R^2}{\alpha^2} \right)  \, , \nonumber
\label{Norm22}
\end{eqnarray}
where $\alpha$ is the Gaussian parameter of the quark wave function.
In the limit where the two baryons overlap ($R \to 0$), the Pauli principle
may impose antisymmetry requirements, due to the existence of identical 
quarks in the interacting baryons, that were not present in a 
hadronic description.
The second diagram in Fig.~\ref{fig1} takes into account the 
exchange of identical quarks 
between the two baryons and would not exist if the quark substructure of 
the baryons had not been considered, the normalization in this case being 
a mere multiplicative factor.
These effects would be prominent in $S$-waves, 
when the two baryons can approach without centrifugal barrier constraints.
Using the asymptotic form
of the Bessel functions, $i_{L+1/2}$, 
in the overlapping region the normalization kernel
of Eq.~\eqref{norm} can be expressed as,
\begin{eqnarray}
{\cal N}_{\Omega^i\Omega^i}^{LS} (R)  
\stackrel[R\to 0]{}{\hbox to 20pt{\rightarrowfill}}
&& 4 \pi \left[ {1 - {\frac{3 R^2}{4 \alpha^2}}} \right]
\frac{1}{1 \cdot 3 \cdots (2L+1)} \, 
\left[ \frac{R^2}{4 \alpha^2} \right]^L \label{norm2}\\
&& \times  \left\lbrace \left[ 3^L - 3 \, C(S) \right]
+  \frac{\left[ 3^{L+2} - 3 \, C(S) \right]}{2(2L+3)} \left[ \frac{R^2}{4 \alpha^2}
\right]^2  
+ \cdots \right\rbrace \, . \nonumber
\end{eqnarray}
$C(S)$ is a flavor-independent spin coefficient given by,
\begin{equation}
C(S) =  \left< \Omega^i (123) \, , \Omega^i (456) ; S \right| P_{36}^{S}
\left| \Omega^i (123) \, ,\Omega^i (456); S \right> \, .
\end{equation}
The spin coefficients for the different $\Omega^i\Omega^i$ states are given 
in Table~\ref{tab2}.
\begin{table}[!htb]
\begin{ruledtabular}
\begin{tabular}{c|cccc}
 $S$ & $0$   &  $1$ &  $2$ &  $3$ \\ \hline
$C(S)$  &   $-\frac{1}{3}$ &   $-\frac{1}{9 }$ & $\frac{1}{3}$ &   $1$ \\
\end{tabular}
\end{ruledtabular}
\caption{$C(S)$ spin coefficients for the $\Omega^i\Omega^i$ states.}
\label{tab2}
\end{table}

It can be seen from Eq.~\eqref{norm2} how in $S$-waves, 
the closer $C(S)$ is to $1/3$ the greater 
the suppression of the normalization of the 
two-baryon wave function at short distances,
generating Pauli repulsion
due to the presence of a quasi-forbidden state~\cite{Oka87}.
For $S$-waves, if $C(S)=\frac{1}{3}$ the normalization of the two-baryon
wave function behaves as $R^{4}$
instead of being a constant, indicating
that {\em Pauli blocking} occurs or, in other words,
a node appears in the relative wave function due to a forbidden
state~\cite{Sai69}. The repulsion due to the presence of
a quasi-forbidden state is much stronger than in other channels where 
the antisymmetrization alone produces the repulsion~\cite{Oka87}.
This is because the existence of a forbidden state 
is a saturation phenomenon that produces a repulsive 
hard-core independent of the dynamics~\cite{Oka81,Oky81}.

There are examples in the literature about the existence 
of quasi-forbibben and forbidden states.
Thus there are, for example, quasi-forbidden channels 
in the $\Sigma N$~\cite{Gar07} and $\Sigma_c N$~\cite{Gar19}
interactions with $(S,I) = (0,1/2)$ 
and $(S,I) = (1,3/2)$, where $C(S) = 8/27$ and $7/27$, 
respectively, giving rise to strong repulsive potentials.
Forbidden states, Pauli blocking, appear, for example,
in the $(S,I)=(1,1)$ $\Delta N$ interaction~\cite{Oka81}. 
It is worth noting that the study of $\pi d$ elastic 
scattering results in a configuration space potential representing 
the $^3S_1 (I=1)$ $\Delta N$ phase shifts which presents 
a strong repulsive core of large radius, greater 
than 1 fm~\cite{Fer89}. Such
repulsion reaches to the limit of the wave function,
much further than in the case of quasi-forbidden states.
There is also experimental evidence about 
such a repulsive $^3S_1 (I=1)$ $\Delta N$ potential
from other processes, such as the $\pi$-nucleus 
inelastic scattering ~\cite{Tak86}. The same is true for 
partial waves of the $\Delta\Delta$ interaction,
see, for example, Fig. 5 of Ref.~\cite{Val97}.
The effect of Pauli blocking 
is quite clear, it generates strongly repulsive phase shifts
that leave no room for the existence of bound states,
see, for example, Fig. 3 of Ref.~\cite{Oky81}.

\section{$J^P=3/2^+$ $\Omega^i\Omega^i \Omega^i$ state.}
We can derive the $^5S_2$ $\Omega^i\Omega^i$ interaction
from a basic interaction between the quarks making use of
the Born-Oppenheimer (BO) approximation. 
The BO method, also known as adiabatic approximation, has been frequently
employed for the study of the nuclear force in terms of the microscopic 
degrees of freedom~\cite{Lib77,Oka84,Val05}. It is based on the assumption 
that quarks move inside the clusters much faster than the clusters 
themselves. Then, one can integrate out the fast degrees of freedom 
assuming a fixed position for the center of each cluster, obtaining in this 
way a local potential depending on the distance between the centers of 
mass of the clusters.
Explicitly, the potential is defined as follows,
\begin{equation}
V_{B^\alpha B^\beta (L S ) \rightarrow B^\alpha B^\beta (L^{\prime} S^{\prime})} (R) =
\xi_{L \,S}^{L^{\prime}\, S^{\prime}} (R) \, - \, \xi_{L \,S}^{L^{\prime}\, S^{\prime}} (\infty) \, ,  \label{Poten1}
\end{equation}
\noindent where
\begin{equation}
\xi_{L \, S}^{L^{\prime}\, S^{\prime}} (R) \, = \, {\frac{{\left
\langle \Psi_{B^\alpha B^\beta}^{L^{\prime}\, S^{\prime}} ({\vec R}) \mid
\sum_{i<j=1}^{6} V_{q_iq_j}({\vec r}_{ij}) \mid \Psi_{B^\alpha B^\beta}^{L \, S} ({\vec R%
}) \right \rangle} }{{\sqrt{\left \langle \Psi_{B^\alpha B^\beta }^{L^{\prime}\,
S^{\prime}} ({\vec R}) \mid \Psi_{B^\alpha B^\beta }^{L^{\prime}\, S^{\prime}} ({%
\vec R}) \right \rangle} \sqrt{\left \langle \Psi_{B^\alpha B^\beta }^{L \, S} ({\vec %
R}) \mid \Psi_{B^\alpha B^\beta }^{L \, S} ({\vec R}) \right \rangle}}}} \, .
\label{Poten2}
\end{equation}
In the last expression the quark coordinates are integrated out keeping $R$
fixed, the resulting interaction being a function of the two-baryon
relative distance.

We derive the interaction making use of the chiral constituent quark model 
of Ref.~\cite{Val05}. This model was proposed in the early 90's in an attempt to
obtain a simultaneous description of the nucleon-nucleon
interaction and the light baryon spectra. 
It was later on generalized to all flavor sectors~\cite{Vij05,Val08}. 
In this model hadrons are described as clusters of three interacting 
massive (constituent) quarks.
The masses of the quarks are generated by the 
dynamical breaking of the original 
$\mathrm{SU}(2)_{L}\otimes \mathrm{SU}(2)_{R}$ chiral symmetry of the QCD 
Lagrangian at a momentum scale of the order of 
$\Lambda_{\rm CSB} = 4\pi f_\pi \sim 1$~GeV, 
where $f_\pi$ is the pion electroweak decay constant. For momenta typically below that 
scale, when using the linear realization of chiral symmetry, light quarks 
interact through potentials generated by the exchange of Goldstone 
bosons. Notice that for the particular case of heavy quarks ($c$ or $b$) 
chiral symmetry is explicitly broken and therefore boson exchanges associated to 
the dynamical breaking of chiral symmetry do not contribute.
Perturbative QCD effects are taken into account through the 
one-gluon-exchange (OGE) potential~\cite{Ruj75}.
Finally, any model imitating QCD should incorporate
confinement. Although it is a very important term from the spectroscopic point of view,
it is negligible for the hadron$-$hadron interaction. Lattice QCD calculations 
suggest a screening effect on the potential when increasing the interquark 
distance~\cite{Bal01}. We refer the reader to Refs.~\cite{Val05,Vij05}
for a detailed description of the model.

Fig. 3 of Ref.~\cite{Val05} shows the different quark diagrams contributing
to the baryon-baryon interaction. The kernel of the different diagrams,
in particular the Coulomb-like term of the OGE, is different from zero at
short distances and independent of the mass of the interacting quarks.
According to Eq.~\eqref{Poten2}, the kernels of the different diagrams 
must be divided by the normalization of the two-baryon wave function 
in Eq.~\eqref{norm}.
Inserting the coefficients from Table~\ref{tab2} into the normalization 
of the two-baryon wave function in Eq.~\eqref{norm2}, 
the $^5S_2$ $\Omega^i\Omega^i$ partial wave satisfies the condition
$C(S)=\frac{1}{3}$, which translates into
the suppression in the normalization of the two-baryon
wave function at short-range not expected at 
baryonic level.~\footnote{Let us also note the presence of the symmetry
repulsive hard-core for $L\neq 0$ in the $^7P_{2,3,4}$ $\Omega^i\Omega^i$ partial 
wave, what is also a
feature of the $\Delta\Delta$ interaction in the $^7P_{2,3,4}(I=3)$ partial
wave~\cite{Val97}. However, in these cases, the effects of Pauli blocking
may be masked by the centrifugal barrier.}
Consequently, a strong repulsive hard core,
independent of the dynamics, is expected in the $^5S_2$ $\Omega^i\Omega^i$
interaction for any flavor due to the underlying 
quark substructure~\cite{Oka81,Oky81}.
It is important to highlight that the $^1S_0$ $\Omega^i\Omega^i$ partial-wave 
does not satisfy the $S-$wave Pauli blocking condition $C(S)=\frac{1}{3}$, see Table~\ref{tab2}, 
so that the short-range normalization behaves as a constant and does not induce any 
repulsion beyond that obtained from the dynamics considered. This result is in 
complete agreement with the conclusions deduced from 
lattice QCD~\cite{Gon18,Lyu21,Mat23}.

\begin{figure}[t]
\vspace*{-2.cm}
\includegraphics[width=.5\columnwidth]{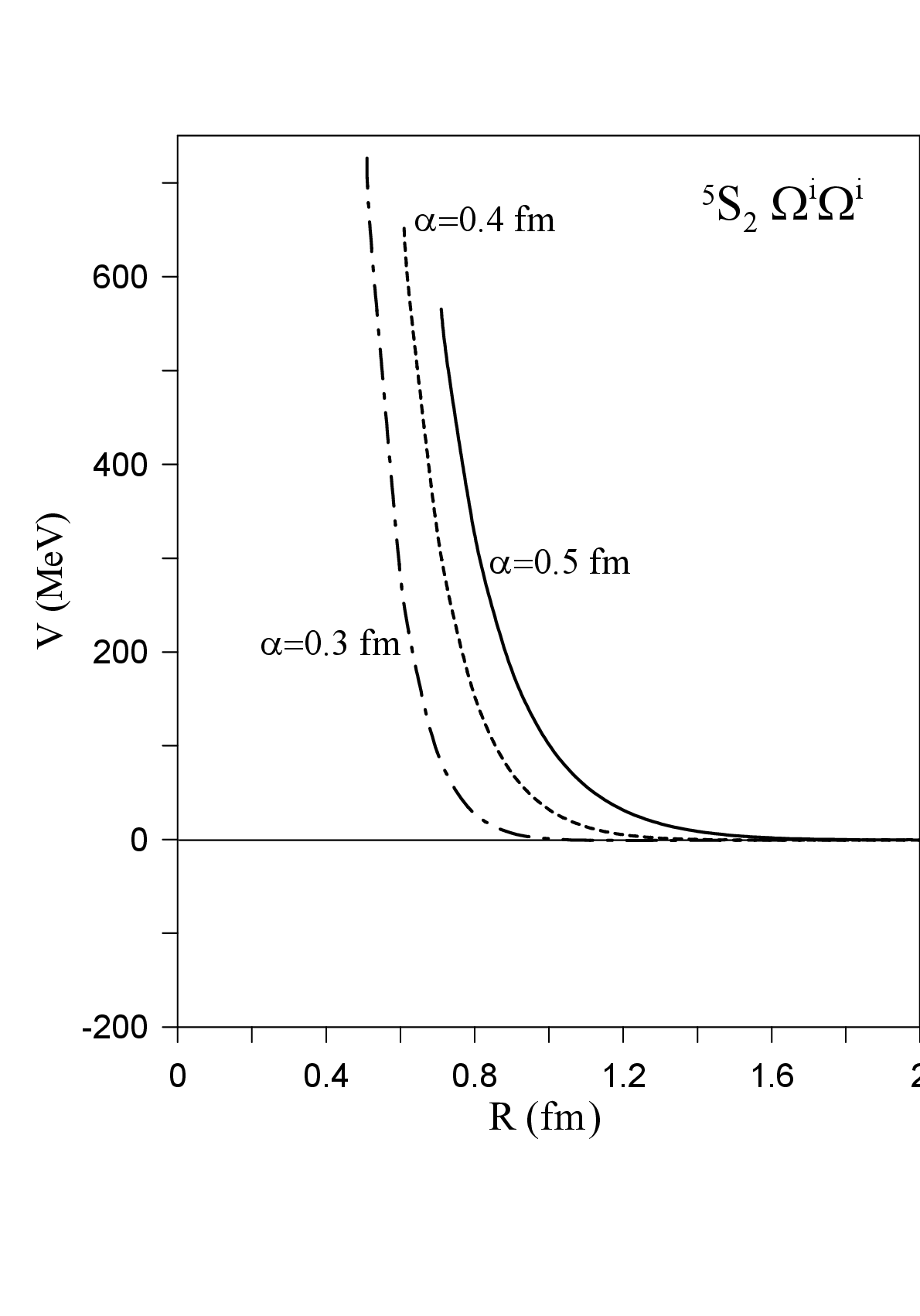}
\vspace*{-1.6cm}
\caption{$^5S_2$ $\Omega^i\Omega^i$
interaction for different values of the Gaussian parameter of the quark
wave function, $\alpha$.}
\label{fig2}
\end{figure}

The $^5S_2$ $\Omega^i\Omega^i$ repulsive interaction is shown in Fig.~\ref{fig2} using the 
parameters and the interacting potential of Ref.~\cite{Gar19}. 
Since the dominant term is the Coulomb-like contribution of the
OGE potential, which is flavor independent, the range of the
repulsive core depends on the Gaussian parameter of the quark
wave function, $\alpha$. We show the potential for three different
values of $\alpha$, representing the typical size of the 
different $\Omega^i$ baryons~\cite{Gal14}. 
As can be seen the potential is repulsive, the repulsive core
going to the limit of the relative wave function.
This result for the $^5S_2$ $\Omega^i\Omega^i$ interaction
was to some extent obvious and expected, because it is impossible 
to accommodate six identical fermions of spin 1/2 of the same flavor 
in the ground state with three colors 
and total spin 2: two of them must necessarily
be in the same quantum state.~\footnote{Note that, 
as in the case of the tetraquarks, the similarities with 
atomic physics are maintained because a dynamics-independent effect due to 
the Fermi-Dirac statistic is responsible, for example, for 
reducing the energy of parahelium compared to that of orthohelium.}
As mentioned above, this result has already been reported in the literature 
for other two-baryon systems showing Pauli blocking~\cite{Oky81}, 
obtaining repulsive cores starting 
about $0.75-0.78$ fm in the case of baryons made of 
lights quarks, i.e., $\alpha\simeq0.5$ fm.
What is really important is that this effect is independent 
of the dynamical model considered as long as interactions 
at the quark level are considered.

It is also important to emphasize that such a 
repulsive hard-core would not appear if 
antisymmetry effects among the baryon constituent 
would have not been considered. This would be 
the case if the interactions were derived without
taking into account the quark substructure of the 
baryons, i.e., by extrapolating 
them either from interactions of other two-body systems or 
from the same system with quantum numbers that were not 
affected by Pauli blocking. This has been clearly illustrated
in Ref.~\cite{Gar24}, showing that a naive ansatz 
of spin independence of the $\Omega^i\Omega^i$ interaction
leads to bound states of hundreds of MeV for the $i=b$ case 
using the lattice QCD interactions of Ref.~\cite{Mat23}.
Even if one assumes that the $^5S_2$ $\Omega^i\Omega^i$ interaction
has a strong repulsive core, but  of the same range 
as that of the $^1S_0$ $\Omega^i\Omega^i$ interaction, 
which would be equivalent to disregarding
the Pauli blocking effects on the $^5S_2$ channel,
the binding energy changes very slowly
leading to bound states of between 250 and 
350 MeV for $i=b$ systems.

In addition to the evidence discussed in the previous section, 
it is important to highlight that there are preliminary studies of lattice QCD
about the $S$-wave scattering of strangeness $-3$ baryons~\cite{Buc12}
concluding a strongly repulsive interaction for $S=2$.
Using lattice configurations with pion mass $m_{\pi} \sim 390$ MeV
and two different volumes, the $\Omega\Omega$ $^1S_0$ channel was 
found to be most likely weakly repulsive or attractive, leading to an extrapolated
scattering length $a^{\Omega\Omega}_{S=0}= 0.16 \pm 0.22$ fm.
However, the states achieved for $S=2$ are at a significantly higher 
energy level than for the $S=0$ case, implying a strongly repulsive 
channel. In fact, it is suggested that none of the extracted states
were actually the $S=2$ ground state, pointing to further
studies at different pion masses approaching the physical
point to gain a better understanding. 

Finally, we have solved the full $\Omega^i\Omega^i\Omega^i$ 
three-body problem by the Faddeev
method~\cite{Gar97} considering the $^1S_0$ and $^5S_2$ partial waves. 
We have used as input the $^1S_0$ $\Omega^i\Omega^i$ potentials 
derived from the lattice QCD calculations~\cite{Gon18,Lyu21,Mat23}
and the $^5S_0$ $\Omega^i\Omega^i$ potentials
of Fig.~\ref{fig2}. The repulsive character of the single channel
calculation due to the Pauli effects in the $^1S_0$ partial wave
are reinforced due to the strong repulsion of the $^5S_2$
$\Omega^i\Omega^i$ interactions due to Pauli blocking effects.
In spite of the attractive character of the $^1S_0$ two-body interactions,
we have not found $J^P=3/2^+$ three-body bound states in any 
flavor sector.

Thus, two different consequences of the Pauli principle
conspire against the possible existence
of $\Omega^i\Omega^i\Omega^i$ bound states. On the one hand,
at the baryon level, the negative recoupling coefficient of 
the spin-0 two-body amplitudes that effectively changes the nature of 
the $^1S_0$ $\Omega^i\Omega^i$ 
two-body interactions from attractive to repulsive.
On the other hand, since the $\Omega^i$'s are spin $3/2$ fermions,
they also contribute spin-2 amplitudes that are free of
Pauli effects at baryonic level. 
The contribution of the spin-2 amplitudes to the
unique $S$-wave fully antisymmetric state made of three $\Omega^i$, 
the $J^P=3/2^+$, is greater than
that of the $^1S_0$ partial wave. However, due to the 
underlying quark substructure, spin-2 amplitudes 
present Pauli blocking which translates into a strong
repulsive interaction. The cooperative combination 
of these two consequences of the Pauli principle, 
at the baryon level and at the quark level,
points to the difficulty of the existence
of $S$-wave $\Omega^i\Omega^i\Omega^i$ bound states 
for any flavor sector.

\section{Summary and outlook.}
In brief, we have analyzed the current question regarding the 
possible existence of $S$-wave 
$\Omega^i\Omega^i\Omega^i$ bound states for any flavor $i=s,c,b$. 
This possibility is suggested by the attractive character of the  
$^1S_0$ $\Omega^i\Omega^i$ interactions derived by lattice QCD simulations,
particularly strong for the heavier flavor $i=b$. We have justified the 
importance of performing a full fledged 
study of the three-body system, including the 
$^5S_2$ $\Omega^i\Omega^i$ partial wave, which has not yet been 
studied by lattice QCD.
We have shown that, as in the three-neutron case, the three-body
recoupling coefficients effectively change the nature 
of the $^1S_0$ $\Omega^i\Omega^i$ two-body interaction 
from attractive to repulsive in the three-body system.
We have also found that the dominant $^5S_2$ partial wave shows Pauli blocking,
a saturation phenomenon independent of the dynamics 
that translates into a strong repulsive interaction. 
It is important to highlight how the $^1S_0$ $\Omega^i\Omega^i$ partial-wave 
does not satisfy the $S-$wave Pauli blocking condition.
As stated above, the cooperative combination of these two consequences of 
the Pauli principle, at the baryon level and at the quark level,
points to the difficulty of the existence of $\Omega^i\Omega^i\Omega^i$ bound states 
despite the attractive interaction -- rather strong in some flavor sectors -- 
predicted for the $\Omega^i\Omega^i$ $^1S_0$ 
partial wave by lattice QCD calculations.

The results of this study suggest an effort by lattice QCD to derive the
interaction in the $^5S_2$ $\Omega^i\Omega^i$ channel and reveal the existence
of a strong dynamics-independent repulsion, thus demonstrating the observable
effects of Pauli blocking due to quark substructure. In addition,  
it would also highlight the caution to be taken when extrapolating interactions 
at the baryon level between different systems or quantum numbers without 
taking into account the observable effects of the quark substructure,
the so-called Pauli blocking.

\section{Acknowledgments.}
The authors acknowledge enlightening correspondence with N.~Mathur and 
M.~Padmanath. This work has been partially funded by COFAA-IPN (M\'exico) and 
by the Spanish Ministerio de Ciencia e Innovaci\'on (MICINN) and the
European Regional Development Fund (ERDF) under 
contracts PID2022-141910NB-I00
and RED2022-134204-E.

\end{document}